\documentclass[%
 reprint,
 amsmath,amssymb,
 aps,
]{revtex4-1}

\usepackage{graphicx}% Include figure files
\usepackage{dcolumn}% Align table columns on decimal point
\usepackage{bm}% bold math
\usepackage{hyperref}
\hypersetup{
    colorlinks=true,
    linkcolor=blue,
    filecolor=blue,      
    urlcolor=black,
    pdftitle={Overleaf Example},
    pdfpagemode=FullScreen,
    }% add hypertext capabilities
\begin{document}

\preprint{APS/123-QED}

\title{A general scheme of differential imaging employing weak measurement}% Force line breaks with \\

\author{Xiong Liu}
\author{An Wang}
\author{Junfan Zhu}
\author{Ling Ye}
\author{Rongchun Ge}
\author{Jinglei Du}
\author{Hong Zhang}%
\email{hongzhang@scu.edu.cn}
\author{Zhiyou Zhang}%
\email{zhangzhiyou@scu.edu.cn}
\affiliation{%
 College of Physics, Sichuan University, Chengdu China, 610064}%

\date{\today}% It is always \today, today,
             %  but any date may be explicitly specified

\begin{abstract}
We propose and experimentally realize a general scheme of differential imaging employing the idea of weak measurement. We show that the weak coupling between the system of interest and a two-level ancilla can introduce a two-beam circuit after an arbitrary pre-selection of the ancilla. By choosing the post-selection orthogonal to the pre-selection measurement, an effective imaging platform based on differential operations is shown achieved. Experimental results on both the Sagnac interferometer and ultra-thin Wollaston prism demonstrate that our imaging scheme successfully yields the boundary information of complex geometric configurations. 
\end{abstract}

%\keywords{Suggested keywords}%Use showkeys class option if keyword
                              %display desired
\maketitle

%\tableofcontents

\section{Introduction}

As a strategy to go beyond the strong projective measurement, weak measurement has attracted wide attention ever since its first proposal for yielding seemly unreasonably large experimental results \cite{aharonov1988result}. To get a large weak value, a post-selected measurement which is almost orthogonal to the pre-selected state of the system is necessary. Aside from the pre- and post-selections operations, a weak coupling between the system of interest and the meter is important, to make sure that the meter would not be seriously suffering from dephasing (when tracing out the system) during the whole process. The exotic properties of weak measurement have made it an important research tool to tackle fundamental problems \cite{kocsis2011observing} and explore novel phenomena of quantum mechanics. Among the ongoing endeavor of applications of weak measurement are precision estimations \cite{brunner2010measuring,starling2010continuous,xu2013phase,qiu2017precision, hosten2008observation,dixon2009ultrasensitive,magana2014amplification,wang2020measurement,xia2020high,liu2021high,fang2019improving,fang2016ultra,chen2021beating,yang2020experimental,yin2021improving}, ultrasensitive sensors \cite{li2017molecular, luo2017precision,zhang2016optical,li2019high,wang2020experimental} and fundamental physics research \cite{ yokota2009direct,lundeen2009experimental,brunner2004direct, lundeen2011direct,pan2019direct,liu2020experimental}, and so on. However, to the best of our knowledge a plausible imaging scheme employing weak measurement has been missing up to now.

Boundaries are the places around which the properties of some physical distributions (like density, permittivity etc.) change dramatically, and consequently is the home of various exotic physical phenomena such as optical spin hall effect \cite{hosten2008observation} and the dissipationless edge transport of the topological insulator \cite{kou2014scale,tokura2019magnetic} to name just a few. So it is important to recognize the boundaries of arbitrarily shaped configuration. Once the boundaries are identified, a decent image of the configuration is obtained. Optical analog differential operation is a physical method of achieving imaging. As a basic mathematical operation, differentiation is usually processed directly in digital. However, in some applications where real-time differential computation is required, such as in medical and satellite applications \cite{pham2000current,holyer1989edge}, analog operation offers a new opportunity to the electronic method. Traditionally, analog differential mathematical computations are realized as analog computers \cite{price1984history,clymer1993mechanical}, but these solutions are not used widely because of their large size and slow response. Optics provides us the opportunity of real-time mathematical operations, and intensive studies have been carried out to shrink the size and speed up the response \cite{silva2014performing,hwang2016optical,zhu2017plasmonic,fang2017grating,dong2018optical,doskolovich2014spatial,youssefi2016analog,pors2015analog,golovastikov2015spatial,ruan2015spatial,guo2018photonic}. Especially, Silva {\it et al}. suitably designed metamaterial blocks to perform a variety of mathematical operations, including differential operation \cite{silva2014performing}.  Recently, Luo {\it et al}. successfully apply spin Hall effect of light to spatial differentiation  \cite{zhu2019generalized,zhu2020optical,xu2020optical,he2020spatial}, and propose more optical differential operation models \cite{zhou2021two,zhou2019optical,xu2021enhanced} with the insight of geometric phase, which will greatly enrich the technology of edge detection.
\begin{figure}[htbp]
\centering\includegraphics[width=8.5cm] {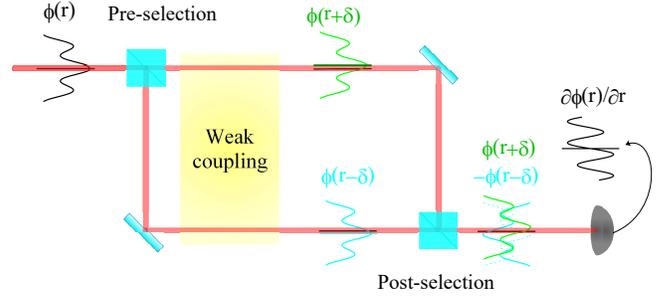}\caption{\label{Fig1} Schematic diagram of weak measurement. Pre-selection, weak coupling and post-selection (perpendicular to the pre-selection) together make up the differential calculator; the output beam is collected at the detector (bottom right).}
\end{figure}

As a continuation of the endeavor of applications of weak measurement, in this paper we report a general and robust experimental scheme of imaging based on differential operation. To be more specific, we will try to image the boundaries of complex geometric distributions with weak measurement based optical analog differential operations. Our basic idea is shown in Fig.\ref{Fig1}, the input wave of arbitrary function $\phi(r)$ passing through the imaging platform becomes the output wave, which is proportional to the standard differential function $\phi^\prime(r)$, and is finally received by the detector.
Pre-selection initializes the whole system by selecting the input of a two-level ancilla, such as polarization, and this serves as the operating handle of the differential calculator consisting of the weak measurement system, to which input function can be loaded. The weak coupling interaction ($\propto |p_1\rangle\langle p_1|-|p_2\rangle\langle p_2|$) introduces path dependent evolution due to the energy splitting between these relevant paths, $P_i$. Finally, in order to achieve the differential operation a proper post-selection is chosen to produce a phase difference of $\Pi$ between the two paths. Because the pre- and post-selections are perpendicular to each other, the intensity of the output wave is very small. Consequently, the input function is greatly compressed, which reduces the detector's load. It should be emphasized the frequency bandwidth here is mainly decided by the response of the optical devices employed which can be replaced easily in our experimental scheme if necessary; as a result, our scheme could be employed for a much wider range of frequency band compared to schemes based on resonant effect.

\section{Theoretical analysis}
Our physical platform consists of a two-level ancilla with observable $\hat{A}$, preselected as $|\Psi_{pre}\rangle=\alpha|0\rangle + \beta|1\rangle$, and the system of interest whose information such as the geometric configuration is encoded in $\phi(r)$. The effective interaction Hamiltonian is given by $H=g\hat{A}k$, where $g$ is the weak coupling strength, and $k$ is the conjugate variable of the spatial coordinate $r$. Of special interest, we have $\hat{A} = |0\rangle\langle 0| -|1\rangle\langle 1|$ with $|0\rangle$, $|1\rangle$ the orthogonal basis of the two level ancilla. The physical distribution function $\phi(r)$ is Fourier transformed into $\tilde{\phi}(k)$ before the interaction is turned on. Due to the interaction between the system and the meter as represented by a unitary operation $\hat{U}=exp(-igk\hat{A})$, the whole system is
\begin{eqnarray}
|\Phi(k)\rangle&\equiv&\langle k|\Phi\rangle
=\hat{U}|\Psi_{pre}\rangle\tilde{\phi}(k)\nonumber\\
&=&(\alpha e^{-igk}|0\rangle+\beta e^{igk}|1\rangle)\tilde{\phi}(k)\label{e1}. 
\end{eqnarray}
By inverse Fourier transform we obtain, $|\Phi(r)\rangle=(\alpha\phi(r-g)|0\rangle+\beta\phi(r+g)|1\rangle)$, which means that after the interaction, there is a split of size $g$ between the $|0\rangle$ and $|1\rangle$ components. Then a special post-selection perpendicular to the pre-selection state, i.e., $|\Psi_{post}\rangle =|\Psi_{pre}^{\perp}\rangle=\mu|0\rangle+\nu|1\rangle$ is chosen, where $|\mu|^2+|\nu|^2=1$, $\mu^{\star}\alpha+\nu^{\star}\beta=0$. Finally, the whole system can be expressed as 
\begin{equation}
\langle\Psi_{pre}^{\perp}|\Phi(r)\rangle=\mu^{\star}\alpha(\phi(r-g)-\phi(r+g))\label{e2}.
\end{equation}
It is worth noting that the post-selection gives a phase difference of $\pi$ between the two paths $|0\rangle$ and $|1\rangle$. In weak measurement, the spatial distance over which the initial field of image displaying a considerable change (like an obvious deviation from a linear increase/decrease) is much larger than the typical distance decided by the coupling strength $g$, so the final output wave function is approximately proportional to the first-order differentiation of the input wave function,
\begin{equation}
\langle \Psi_{pre}^{\perp}|\Phi(r)\rangle\simeq-2\mu^{\star}\alpha g\phi^{\prime}(r)\label{e3}.
\end{equation}

However, if the post-selection is introduced without the coordinate in Eq.~(\ref{e1}) being transformed from k-space to r-space, the final output wave function is
\begin{equation}
\langle \Psi_{pre}^{\perp}|\Phi(k)\rangle\simeq-i2\mu^{\star}\alpha gk\tilde{\phi}(k)\label{e4}.
\end{equation}
It is clear that the output beam is proportional to the function $k\tilde{\phi}(k)$, which is not the first-order differentiation of the input function $\tilde{\phi}^{\prime}(k)$ most of the time except for Gaussian beam whose first derivative is $k\tilde{\phi}(k)$. From Eqs.(\ref{e3}) and (\ref{e4}), it can be seen that the differential function appears only in the same space as the small split $g$. This observation indicates that the weak coupling parameter and the input distribution should be in the same space to achieve the desired differential operation.

Although the theory shown is for a single photon, the same idea applies to a macroscopic beam which is the case in our experiments in the next section. In the following, we will give a brief description of the parallel framework for electromagnetic waves at the classical limit.

Considering a plane wave, we define the $0-$ and $1-$ as the two paths of evolution. Under the paraxial approximation, the incident and outgoing beams have the electric fields
\begin{equation}
E_i=\int[u_0^i\widetilde{E}_0^i(k)+u_1^i\widetilde{E}_1^i(k)]e^{ikr}dk,\label{e5} 
\end{equation}
\begin{equation}
E_o=\int[u_0^o\widetilde{E}_0^o(k)+u_1^o\widetilde{E}_1^o(k)]e^{ikr}dk,\label{e6} 
\end{equation}
where $u_{0(1)}^{i}$ and $u_{0(1)}^{o}$ correspond to the path-dependent incident and outgoing beams, respectively. To calculate the r-space spectral transfer function, we decompose the fields $E_i$ and $E_o$ into plane waves. The pre- and post-selected states are described in matrix form as $V_{pre}=\left(\begin{matrix}\alpha \\\beta\\\end{matrix} \right)$ and $V_{post}=\left(\begin{matrix}\mu \\\nu\\\end{matrix} \right)$. Thus, the Fourier spectrum of the outgoing beam can be obtained as $ \left(\begin{matrix}\widetilde{E}_0^o \\\widetilde{E}_1^o\\\end{matrix} \right)=V_{post}^{\star}UV_{pre}\left(\begin{matrix}\widetilde{E}_0^i \\\widetilde{E}_1^i\\\end{matrix} \right)$
where $U$ is the evolution operator incurred by the weak coupling, which can be evaluated through the geometric phase of the $0-$ and $1-$ paths under the paraxial approximation, as $U=\left(\begin{matrix}\exp(-i \theta_g/2) & 0 \\0 & \exp(i\theta_g/2)\\\end{matrix} \right)$.
Here $\theta_g=2gk$ is the geometric phase during the weak coupling. Therefore, the evolution in k-space could be effectively described by $H(k)\equiv\widetilde{E}^o(k)/\widetilde{E}^i(k)=V_{post}^{\star}UV_{pre}$. Then, the spectral transfer function is obtained as
\begin{equation}
H(k)=\mu^{\star}\alpha (e^{-igk}-e^{igk})\label{e7}.
\end{equation}
For $g|k|\ll1$, we have $H(k)\simeq -i2\mu^{\star}\alpha gk$. As a result, the outgoing field in k-space is $E^o(k)=-2\mu^{\star}\alpha g k E^i(k)$. It should be noticed that the outgoing field is $g k$ times the input field; in terms of light intensity, the output light is only $(g k)^2$ times the input light. In other words, the data of the input function is compressed to as small as $(g k)^2$ times. And in r-space, the outgoing field reads
\begin{equation}
E^o(r)=-2\mu^{\star}\alpha g\dfrac{\partial E^i(r)}{\partial r}\label{e8},
\end{equation}
which agrees with Eq.~(\ref{e3}). In summary, the theory shows that the differential operation can be realized via arbitrary two-level weak measurement system.

\section{Experimental observation}
In this section, we show two experimental realizations of our weak measurement based imaging proposal: (1) Sagnac interferometer based differential imaging platform, in which the Sagnac interferometer realizes the differential shear distance by a small deflection of the rotating mirror(RM), and at the output, the spatial mathematical differential operation is achieved; (2) Wollaston prism based differential imaging platform, in which an ultrathin Wollaston prism performs the differential shear distance ($\delta$), and the mathematical differential operation in the spatial domain is realized along with the whole weak measurement system.    

\subsection{Sagnac interferometer based differential imaging platform}
\begin{figure}[htbp]
\centering\includegraphics[width=8.5cm] {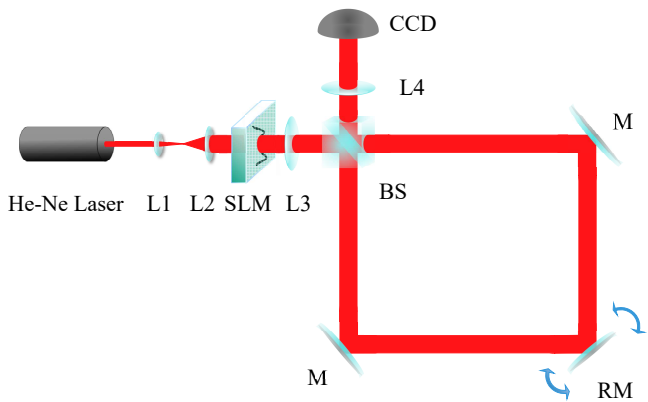}\caption{\label{Fig2} Experimental setup: the light source is He-Ne laser (wave length $\lambda_0=632.8\ {\rm nm}$); L1 and L2, beam expander, spatial light modulator (SLM); L3 and L4, lenses with focal lengths $250\ {\rm mm}$; beam splitter (BS); mirror (M); rotating mirror (RM); CCD, Thorlabs BC106N-VIS.}
\end{figure}

In our Sagnac interferometer based differential imaging platform, the system consists of three cascaded subblocks: (i) a Fourier transform lens (L3), (ii) Sagnac interferometer, realizing three steps of weak measurement, and (iii) an inverse Fourier transform lens (L4) as shown in Fig.\ref{Fig2}.

The laser from the He-Ne laser is expanded by a beam expander at the first step; it then obtains an image with specific spatial distribution as passing through the spatial light modulator (SLM), denoted as red $|\psi_i\rangle=\int dr\phi_i(r)|r\rangle$, where $r$ is the spatial coordinate. In the paraxial approximation, L3 accomplishes the Fourier transform of the image into the momentum space, i.e., $|\psi_i\rangle=\int dk\widetilde{\phi}_i(k)|k\rangle$. When the light passes through the beam splitter (BS), it splits into two paths, propagating along the clockwise and counterclockwise directions, respectively. Here, the path freedom of the light beam constitutes a two-level system ($|\circlearrowright\rangle$ and $|\circlearrowleft\rangle$). In other words, the light is pre-selected as the state of the superposition of two paths, i.e., $|R_{pre}\rangle=\dfrac{1}{\sqrt{2}}(|\circlearrowright\rangle+i|\circlearrowleft\rangle)$, where $i$ is the phase $\pi/2$ of the light induced as the counterclockwise path reflects at the BS. And the whole initial state is $|\Psi_i\rangle=|\psi_i\rangle|R_{pre}\rangle$. Rotating mirror (RM) is a mirror that can be rotated by a small angle $\beta$, which tilts the lights in the clockwise and the counterclockwise paths slightly in opposite directions, $\Delta\theta=4\beta$, and this tilt ($\Delta\theta$) leads to a small displacement of the transverse momentum of the beam ($\Delta k_x=k_0\Delta\theta$). The observable of paths and the continuous transverse tilt together form the weak interaction, described by $\hat{U}=e^{-i k_x\hat{A}\delta/2}$ with $\delta=l\Delta\theta$, where $\hat{A}=|\circlearrowright\rangle\langle\circlearrowright|-|\circlearrowleft\rangle\langle\circlearrowleft|$ is the observable of paths; $l$ is the distance from RM to CCD \cite{dixon2009ultrasensitive}, and $k_x$ is 
the momentum in x-direction for RM rotating in the horizontal plane only. When the beams pass through the BS again, the post-selection applies, $|R_{post}\rangle =\dfrac{1}{\sqrt{2}}(|\circlearrowright\rangle-i|\circlearrowleft\rangle)$, with the final state of light given by
\begin{eqnarray}
|\Psi_f\rangle&=&\langle R_{post}|\hat{U}|\Psi_i\rangle\nonumber\\
 &=&\dfrac{1}{2}\int dk\widetilde{\phi}_i(k)(e^{-ik_x\delta/2}-e^{ik_x\delta/2})|k\rangle\label{e9}.
\end{eqnarray}
After the inverse Fourier transform by L4, the outgoing state is
\begin{equation}
|\Psi_o\rangle= -\dfrac{1}{2}\delta\int dr\dfrac{\partial\phi_i(x)}{\partial r}|r\rangle\label{e10}.
\end{equation}
Finally, our CCD measures the light, which is the first derivative in the x-direction of the specific spatial distribution image $\phi_i(r)$.

\begin{figure}[htbp]
\centering\includegraphics[width=8.5cm] {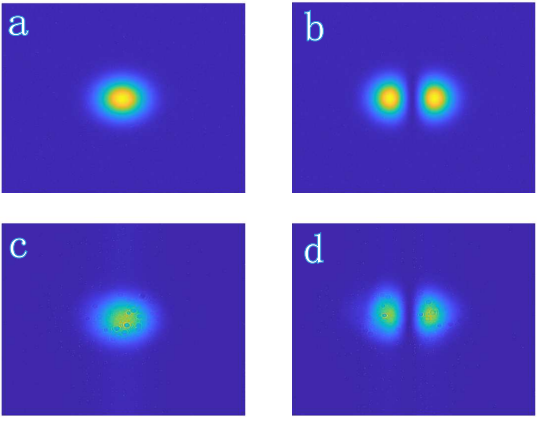}\caption{\label{Fig3} Gaussian distributions and the differential images of them. (a) and (c) are simulated and experimental Gaussian distributions, respectively. (b) and (d) are the corresponding differential images in x-direction.}
\end{figure}
To illustrate the spatial differentiation effect, we measure the outgoing field distribution under a Gaussian beam illumination firstly. Figures \ref{Fig3}(c) and \ref{Fig3}(d) show the measured intensity profiles for the incident and outgoing beams, respectively. To quantitatively illustrate the performance of spatial differentiation, the incident beam is numerically fitted with a Gaussian profile, and its differential function in x-direction is calculated numerically, as shown in Figs.\ref{Fig3}(a) and \ref{Fig3}(b), respectively. The experimental outgoing fields show a good agreement with the ideal spatial differentiation.  

\begin{figure}[htbp]
\centering\includegraphics[width=8.5cm] {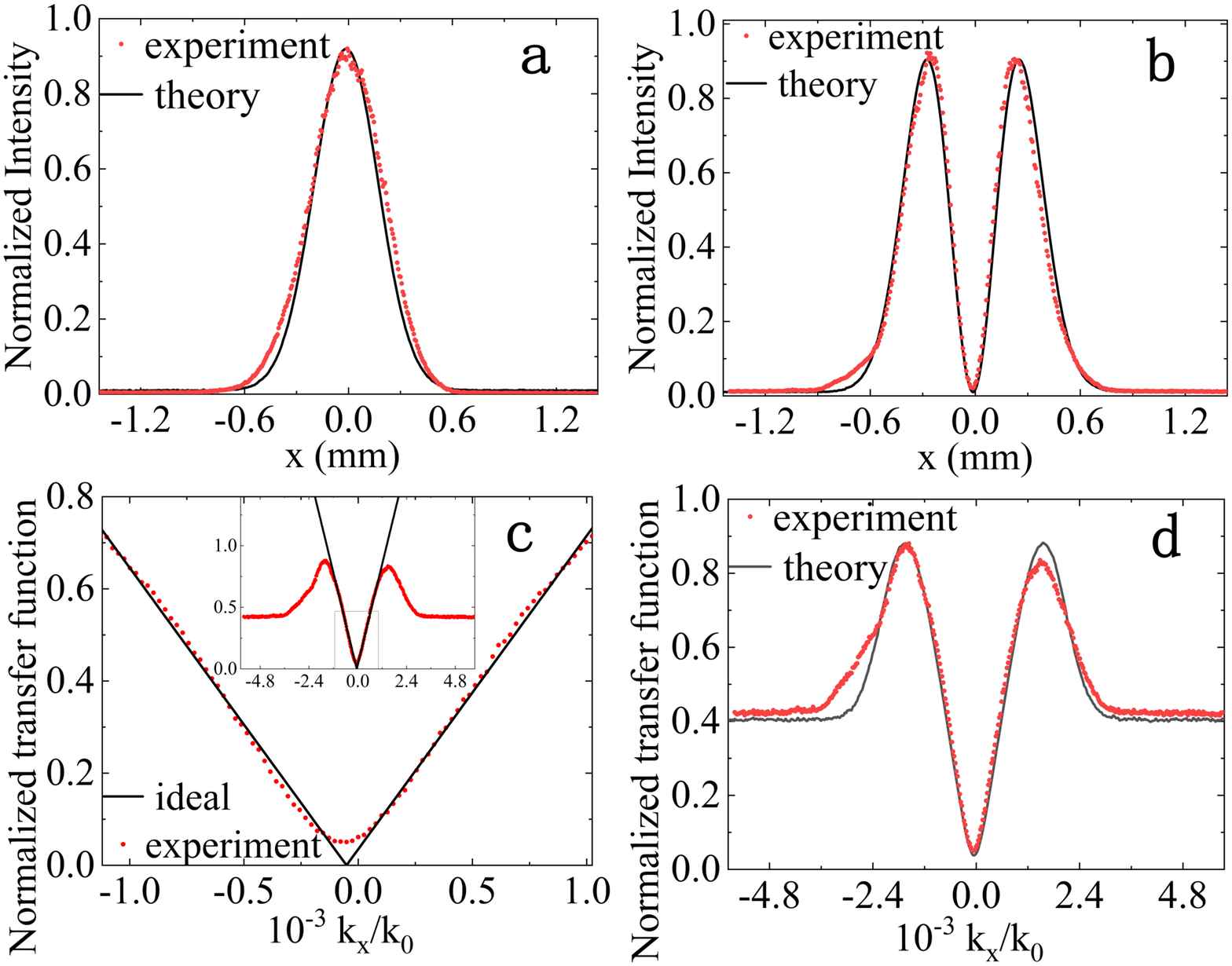}\caption{\label{Fig4} Measurement of the spatial spectral transfer function. (a) Gaussian beam for $k_y=0$. (b) The first derivative of Gaussian beam in x-direction for $k_y=0$. (c) Spatial spectral transfer function for $k_y=0$. (d) Spatial} spectral transfer function for $k_y=0$ (consider a tiny random light noise (0.01\%).). Experimental results are shown in red dots, and theoretical results are in black lines.
\end{figure}
Then, the spatial transfer function of the Sagnac interferometer based  differential imaging platform is calculated based on experiments. From the relation of the incident and outgoing spatial spectra, we have the spatial transfer function, $H(k)=E^o(k)/E^i(k)$. The outgoing field distribution under a Gaussian beam can be measured with the device shown in Fig.\ref{Fig2}, i.e., outgoing spatial spectrum is at the back focal plane of L4. Then we block arbitrarily one of the paths, and the incident spatial spectrum is at the back focal plane of L4. Since CCD can only receive the intensity distribution of light in spatial space, we calculate the electric field with the equation of $|E(r)|=\sqrt{I(r)}$. With the help of L4, the spatial coordinate $r$ is related to the momentum $k$, $r=f k/k_0$, where $f$ is the focal distance of L4, so we obtain $E(k)=E(r)k_0/f$. Finally the spatial transfer function can be calculated as
\begin{equation}
H(k)=\sqrt{I_o(k)}/\sqrt{I_i(k)}=\sqrt{I_o(r)}/\sqrt{I_i(r)}\label{e11}, 
\end{equation}
and the theoretical transfer function is obtained by $H(k)=E_G^\prime(k)/E_G(k)$, where $E_G(k)$ is the electric field of Gaussian function. The results are shown in Fig.\ref{Fig4}: it is clear from Fig.\ref{Fig4}(c), the experimental transfer function matches that of the theory well in the central region, but not in other regions, as a result of the pervasive noise. Even though it is tiny (\textless 0.01\%), the noise could seriously affect the experimental results since the amplitude of Gaussian beam is small away from the central region. To roughly account the ubiquitous noise residing in the laser beam, such as the randomly distributed speckles, we add a uniformlly distributed random term, i.e., $H(k)=[E_G^\prime(k)+c\times  rand(k)]/[E_G(k)+c\times rand(k)]$,  where $c$ is less than 0.01\% of the amplitude of the otherwise ideal Gaussian beam; then the theoretical and experimental results agree well with each other in all regions, as shown in Fig.\ref{Fig4}(d).

\begin{figure}[htbp]
\centering\includegraphics[width=8.5cm] {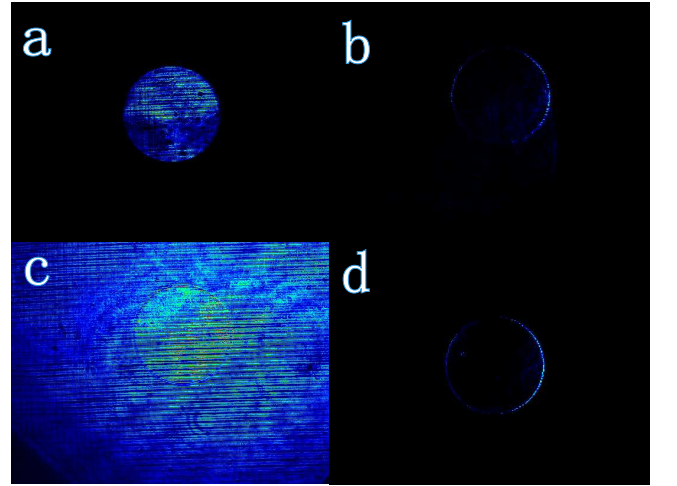}\caption{\label{Fig5} Disk distributions and the differential images of them. (a) and (c), incident images of disk distributions generated with amplitude and phase modulation, respectively. (b) and (d), the corresponding differential images in x-direction.}
\end{figure}
According to the differential imaging theory above, the differential operation can be performed in any direction, but here we only measure the derivative of the incident fields in x-direction as our RM in the Sagnac interferometer can only rotate in horizontal plane (x-direction). Figure \ref{Fig5}(a) shows the incident image of a disk generated with amplitude modulation by using SLM, where the inside region of the circle is filled with light, while the outside is dark. Figure \ref{Fig5}(b) shows the measured outgoing intensity distribution. It clearly exhibits the outline of the disk with spatial differentiation. In Fig.\ref{Fig5}(b), since the differentiation is along x-direction, the edges perpendicular to x-direction are most visible and that of x-direction are invisible. Furthermore, the edge of input beam can be detected as long as it is not completely along x-direction. 

Since the differentiation operates on the electric field rather than on the intensity, the platform can be used to detect the edge of the incident field either in the phase or the amplitude distribution. To show such an effect, we also generate incident field with phase modulation, as shown in Fig.\ref{Fig5}(c), the disk is filled with light both inside and outside. Again, the outgoing light clearly exhibits the edge of the disk in x-direction as shown in Fig.\ref{Fig5}(d). It should be noticed that the intensity of the right side of the circle is stronger than that of the left side, as shown in Figs.\ref{Fig5}(b) and \ref{Fig5}(d). The reason is that the light is not equally divided into two beams by the BS employed in the experiment. 

\begin{figure}[htbp]
\centering\includegraphics[width=8.5cm] {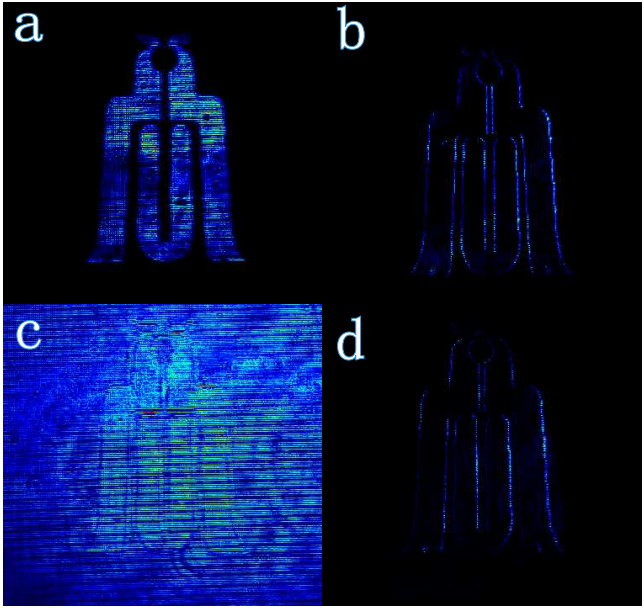}\caption{\label{Fig6} Bell distributions and the differential images of them. (a) and (c), bell distributions generated with amplitude and phase modulation, respectively. (b) and (d), the corresponding differential images in x-direction.}
\end{figure}
In order to show that general graphs can be differentially imaged by the platform, bell-shaped incident fields are generated. It can be seen from Fig.\ref{Fig6} that the platform can differentiate the incident field well no matter it is amplitude or phase distribution. Especially, for the phase distribution with a bright mess as shown in Fig.\ref{Fig6}(c), the edge profile of the bell in x-direction is surprisingly clear after the operation of the differential platform, as shown in Fig.\ref{Fig6}(d). 

It can be seen from Figs.\ref{Fig5} and \ref{Fig6} that the output differential images carry only the edge information of the input images, and there is no light elsewhere; that is to say, the amount of data obtained by CCD is greatly compressed, as a consequence, the detector’s load is sharply decreased, which prevents the CCD from being oversaturated by too strong light.

\subsection{Wollaston prism based differential imaging platform}
\begin{figure}[htbp]
\centering\includegraphics[width=8.5cm] {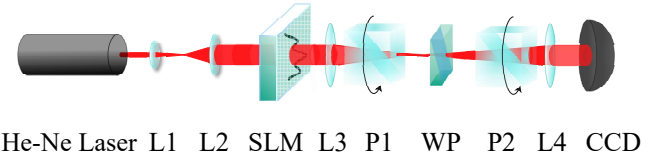}\caption{\label{Fig7} Experimental setup: The light source is He-Ne laser (wave length $\lambda_0=632.8\ {\rm nm}$). L1 and L2, beam expander, SLM, spatial light modulator; P1 and P2, polarizers; L3 and L4, lenses with focal lengths $250\ {\rm mm}$; WP, an ultrathin Wollaston prism; CCD, Thorlabs BC106N-VIS.}
\end{figure}

Our system of Wollaston prism based differential imaging platform consists of five cascaded subblocks: (i) a Fourier transform lens (L3), (ii) a pre-selection (P1), (iii) a weak coupling (WP), (iv) a post-selection (P2), and (v) an inverse Fourier transform lens (L4), as shown in Fig.\ref{Fig7}. 

\begin{figure}[htbp]
\centering\includegraphics[width=8.5cm] {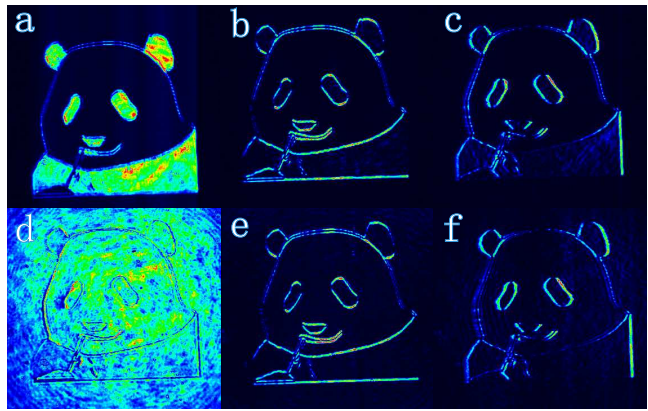}\caption{\label{Fig8}Panda distributions and the differential images of them. (a) and (d), panda distributions generated with amplitude and phase modulation, respectively. (b) and (e), the corresponding differential images in y-direction. (c) and (f), the corresponding differential images in x-direction.}
\end{figure}

As the laser passes through L3, the state of light is written as $|\psi_i\rangle=\int dk\widetilde{\phi}_i(k)|k\rangle$. P1 plays the role of pre-selection and pre-selects the light as
\begin{equation}
|\Psi_i\rangle=|\psi_{pre}\rangle|\psi_i\rangle\label{e12},
\end{equation}
where $|\psi_{pre}\rangle=\dfrac{1}{\sqrt{2}}(|o\rangle+|e\rangle)$ is the pre-selected polarization state of light, $|o\rangle$ and $|e\rangle$ indicate ordinary and extraordinary light, respectively. WP plays the role of weak coupling; as a result of the effect of the ultra-thin WP, ordinary and extraordinary light will split slightly with each other. Suppose the splitting distance is $\delta$ (about 20 $\mu$m), then the weak interaction can be written as $|\Psi_{int\rangle}=\hat{U}|\Psi_i\rangle$, where $\hat{U}=\exp({-ik_d\hat{A}\delta/2})$ is the evolution operator; $\hat{A}=|o\rangle\langle o|-|e\rangle\langle e|$ is the observable operator and $k_d$ is the momentum in d-direction (d-direction means arbitrary direction, in our experiments here, d- is y- or x-). P2 is a post-selected polarizer perpendicular to P1, the state of P2 is described as $|\psi_{post}\rangle=\dfrac{1}{\sqrt{2}}(|o\rangle-|e\rangle)$. And then the light is inverse Fourier transformed into the coordinate space by L4, so the state of outgoing light measured by CCD is
\begin{equation}
|\Psi_o\rangle=-\dfrac{1}{2}\delta\int dr\dfrac{\partial\phi_i(r)}{\partial r_d}|r\rangle\label{e13}.
\end{equation}
From Eq.~(\ref{e13}), the outgoing light is the first derivative in the d-direction of the specific spatial distribution image $\phi_i(r)$.

Here we choose two directions to illustrate that weak measurement based differential schemes can operate on the field in any direction. The relevant results are shown in Fig.\ref{Fig8}, both the edge information of phase and amplitude distributions of the incident fields (Panda) can be obtained. The incident amplitude spatial spectrum is measured by removing WP and rotating the polarizer P2 to the same direction as P1, as shown in Fig.\ref{Fig8}(a). Figures \ref{Fig8}(b) and \ref{Fig8}(c) show the measured outgoing intensity distributions being differentiated in y- and x-directions, respectively. They clearly exhibit the outlines of Panda with spatial differentiation. The phase distributions are shown in Figs.\ref{Fig8}(d)-(f). More details are shown in appendix.

\section{Conclusion}
In conclusion, we have proposed a general scheme of imaging employing the idea of weak measurement and listed two differential imaging platforms. Also, we have experimentally demonstrated the generality of spatial differentiation of the weak measurement differential system. Because the pre- and post-selections are perpendicular to each other, the signal-to-noise ratio can be improved effectively while the background noise reduced. Moreover, with no analog-to-digital conversion or other systematic delays, here mathematical operations have already got processed as the electromagnetic signals propagate through the weak measurement system. As a result, the data of the input image is greatly compressed before being received by CCD. Such designs prevent the detector from saturating itself with too strong light and thus missing the signal. 

\begin{acknowledgments}
This work is supported by the National Natural Science Foundation of China (11674234), the Science Specialty Program of Sichuan University (2020SCUNL210).

The authors thank Jiang Liu for the Panda picture.
\end{acknowledgments}

\section{appendix: The spatial transfer function of the Wollaston prism based differential imaging platform and its operation on circle distributions}

\begin{figure}[htbp]
\centering\includegraphics[width=8.5cm] {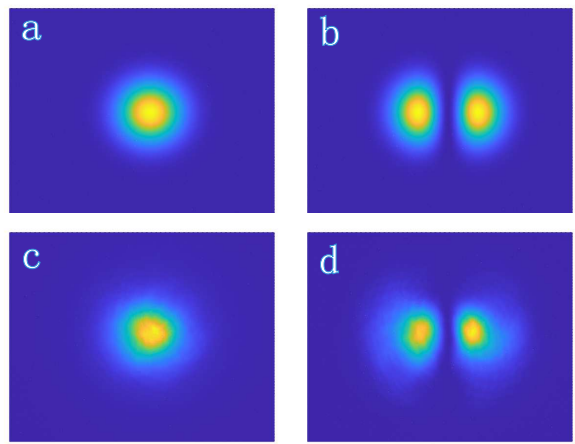}\caption{\label{FigA9}Gaussian distributions and the differential images of them. (a) and (c) are the simulated and experimental Gaussian distributions, respectively. (b) and (d) are the corresponding differential images in x-direction.}
\end{figure}

\begin{figure}[htbp]
\centering\includegraphics[width=6.5cm] {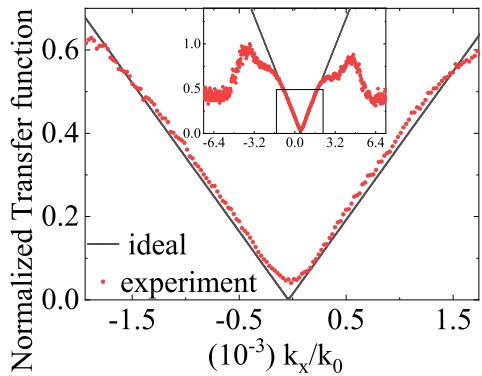}\caption{\label{FigA10} Measurement of the spatial spectral transfer function for $k_y=0$. Experimental results are shown in red dots, and theoretical results are in black lines.}
\end{figure}

The spatial transfer function of the Wollaston prism based differential imaging platform is calculated similarly as the Sagnac interferometer based differential imaging platform. Firstly, we measure the field distribution under a Gaussian beam illumination at the back focal plane of L4. Figures \ref{FigA9}(c) and \ref{FigA9}(d) show the measured intensity profiles for the incident and outgoing beams, respectively. Then, the incident beam is numerically fitted with a Gaussian profile, as shown in Figs.\ref{FigA9}(a) and \ref{FigA9}(b). The experimental outgoing fields show a good agreement with the ideal spatial differentiation. Finally, the spatial transfer function is calculated as $H(k)=\sqrt{I_o(k)}/\sqrt{I_i(k)}$. As shown in Fig.\ref{FigA10}, the results of experiment match the theory well in the central region, but not in other regions, and the reason is the same as the Sagnac interferometer based differential imaging platform. 

\begin{figure}[htbp]
\centering\includegraphics[width=8.5cm] {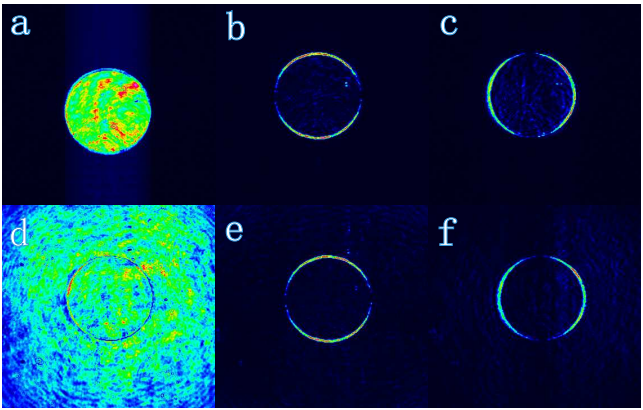}\caption{\label{FigA11} Disk distributions and the differential images of them. (a) and (d) disk distributions generated with amplitude and phase modulation, respectively. (b) and (e), the differential images of disks in y-direction. (c) and (f), the differential images of disks in x-direction.}
\end{figure}

Disk input function is also operated here to more clearly show the characteristics of differential image in different directions. Figure \ref{FigA11}(a) shows the incident image of a disk of light generated with amplitude modulation by using the SLM, where the inside of the disk is filled with light, while the outside is dark. Figures \ref{FigA11}(b) and \ref{FigA11}(c) show the measured outgoing intensities of y- and x-directions, respectively. It clearly exhibits the outline of the disk with spatial differentiation. In Fig.\ref{FigA11}(b), since the differentiation is along the y-direction, the edges perpendicular to the y-direction are most visible and that of y-direction are invisible. Furthermore, as long as the edge is not completely along the y-direction, it can be detected in the outgoing beam. The same thing happens in the x-direction as shown in Fig.\ref{FigA11}(c), the edges perpendicular to the x-direction are most visible and that of x-direction are invisible. 

As the differentiation operates on the electric field rather than on the intensity, the device can be used to detect an edge either in the phase or the amplitude distribution of the incident field. To show such an effect, we also generate incident field with phase modulation. Figure \ref{FigA11}(d) shows the incident image of a circle of light generated with phase modulation, the circle is filled with light both inside and outside. Again, the outgoing light clearly exhibits the edge of the circle in y-direction or x-direction as shown in Figs.\ref{FigA11}(e) and \ref{FigA11}(f).

% The \nocite command causes all entries in a bibliography to be printed out
% whether or not they are actually referenced in the text. This is appropriate
% for the sample file to show the different styles of references, but authors
% most likely will not want to use it.
\nocite{*}

%\bibliography{apssamp}% Produces the bibliography via BibTeX.

\end{document}